\documentclass[twocolumn,aps,showpacs,preprintnumbers,amsmath,amssymb, superscriptaddress]{revtex4-1}
\pdfoutput=1

\usepackage{bm}
\usepackage{graphicx}
\usepackage{color}

\renewcommand{\vec}{\mathbf}

\begin{document}

\title{Comparative Raman study of Weyl semimetals TaAs, NbAs, TaP and NbP}

\author{H. W. Liu}\email{hliu@iphy.ac.cn}
\affiliation{Beijing National Laboratory for Condensed Matter Physics, and Institute of Physics, Chinese Academy of Sciences, Beijing 100190, China}
\author{P. Richard}\email{p.richard@iphy.ac.cn}
\affiliation{Beijing National Laboratory for Condensed Matter Physics, and Institute of Physics, Chinese Academy of Sciences, Beijing 100190, China}
\affiliation{Collaborative Innovation Center of Quantum Matter, Beijing, China}
\author{L. X. Zhao}
\affiliation{Beijing National Laboratory for Condensed Matter Physics, and Institute of Physics, Chinese Academy of Sciences, Beijing 100190, China}
\author{G.-F. Chen}
\affiliation{Beijing National Laboratory for Condensed Matter Physics, and Institute of Physics, Chinese Academy of Sciences, Beijing 100190, China}
\affiliation{Collaborative Innovation Center of Quantum Matter, Beijing, China}
\author{H. Ding}
\affiliation{Beijing National Laboratory for Condensed Matter Physics, and Institute of Physics, Chinese Academy of Sciences, Beijing 100190, China}
\affiliation{Collaborative Innovation Center of Quantum Matter, Beijing, China}

\date{\today}

\begin{abstract}
We report a comparative polarized Raman study of Weyl semimetals TaAs, NbAs, TaP and NbP. The evolution of the phonon frequencies with the sample composition allows us to determine experimentally which atoms are mainly involved for each vibration mode. Our results confirm previous first-principles calculations indicating that the A$_1$, B$_1(2)$, E$(2)$ and E$(3)$ modes involve mainly the As(P) atoms, the B$_1(1)$ mode is mainly related to Ta(Nb) atoms, and the E$(1)$ mode involves both kinds of atoms. By comparing the energy of the different modes, we establish that the B$_1(1)$, B$_1(2)$, E$(2)$ and E$(3)$ become harder with increasing chemical pressure. This behaviour differs from our observation on the A$_1$ mode, which decreases in energy, in contrast to its behaviour under external pressure. 
\end{abstract}



\maketitle



\section{Introduction}
The prediction \cite{X_Wan_PRB83,Balents_Physics2011,G_Xu_PRL107} and recent experimental discovery \cite{Lv_BQ_PRX5,SY_Xu_Science2015,Yang_LX_NPHYS2015,Lv_BQ_nphys11,Yang_LX_NPHYS2015} of Weyl fermions in condensed matter systems, in particular in the so-called Weyl semimetals of the $AX$ ($A$ = Ta, Nb; $X$~=~As, P) family \cite{H_Weng_PRX5,Huang_NCOMM}, have generated a wide enthusiasm in the scientific community that compares with interest for the topological insulators. The simple $AX$ binary compounds open new possibilities for practical applications. Unlike the topological insulators, for which the exotic electronic properties of interest, though dependent on the bulk, occur at the surface, the Weyl nodes in the Weyl semimetals are a property of the bulk electronic structure itself. In addition to the charge sign and the spin of the electronic carriers, the Weyl semimetals introduce chirality as an appealing parameter for applications, and the chiral anomaly in the Weyl semimetals leads to a large negative magneto-resistance \cite{X_Huang_PRX5,Ghimire_JPCM27,Shekhar_nphys11,C_Zhang_PRB92}. However, the properties associated with the Weyl nodes vary significantly across the $AX$ series \cite{CC_Lee_PRB92,Y_Sun_PRB92,ZK_Liu_nmat15}. In order to understand how tuning the electronic properties of the Weyl semimetals for making devices based on powder, single-crystals or films of these compounds, it is necessary to characterize the crystal structure and the lattice dynamics of real samples. 

Here we report a comparative Raman study of $AX$ ($A$ = Ta, Nb; $X$~=~As, P). For each of these materials we observe all the optical modes at the Brillouin zone centre, which consist in one A$_1$ mode, two B$_1$ modes and three E modes. At the first order, we show that the phonon mode energies vary with the  inverse of the square-root of the mass of the atoms mainly involved in the vibrations, which allows us to confirm experimentally the mode assignments in previous first-principles calculations on TaAs \cite{HW_LiuPRB92}. Our comparative study allows us to show that most modes become harder with increasing chemical pressure. In contrast, the energy of the A$_1$ mode becomes smaller with increasing chemical pressure, which differs from its behaviour under external pressure \cite{Y_Zhou_pressure}, thus indicating that chemical pressure and external pressure act differently on the system.

\section{Experiment}

\begin{figure*}[!t]
\begin{center}
\includegraphics[width=0.85\textwidth]{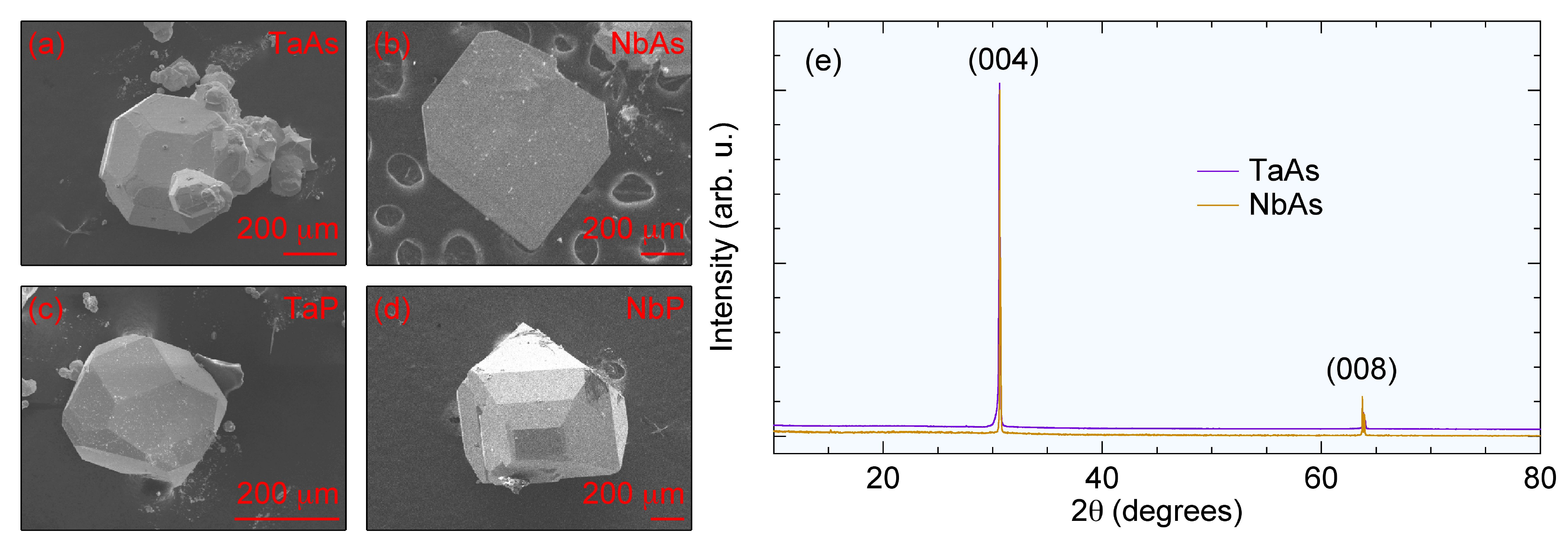}
\end{center}
\caption{\label{SEM_XRD}(Colour online). (a)-(d) SEM images of single crystals of TaAs, NbAs, TaP and NbP, respectively. (e) XRD 2$\theta$ plots of TaAs and NbAs samples similar to the ones used in our study. The curve for TaAs has been shifted up slightly for better visualization.}
\end{figure*}

The single crystals of TaAs, NbAs, TaP and NbP used in this study were grown by chemical vapor transport. In Figs. \ref{SEM_XRD}(a) to \ref{SEM_XRD}(d) we show scanning electron microscopy (SEM) images of typical single crystals recorded with a Hitachi S-4800 microscope. The samples show well-defined surfaces, suggesting good crystallinity. Larger crystals of TaAs and NbAs were also characterized by X-ray diffraction (XRD) using the K$_{\alpha}$ line of a Cu source to determine the crystal orientation. As shown in Figs. \ref{SEM_XRD}(e), only the (004) and (008) peaks are detected, confirming the sample quality and sample orientation. Freshly prepared platelike samples with typical size of $0.4\times 0.4\times 0.08$ mm$^3$ were used for Raman scattering measurements at room temperature. The measurements were performed with the 514.5~nm and 488.0~nm excitations of an Ar-Kr laser focussed on flat sample surface regions with a 100$\times$ objective mounted in a back-scattering micro-Raman configuration. The power at the sample was smaller than 0.4~mW. The signal was analysed by a Horiba Jobin Yvon T64000 spectrometer equipped with a nitrogen-cooled CCD camera. 

All the materials studied have the same non-centrosymmetric structure corresponding to space group I41md ($C_{4v}^{11}$, group no. 109), with 4 atoms in one unit cell. An analysis in terms of the irreducible representations of this group shows that the vibration modes of this system decompose \cite{Comarou_Bilbao} into [A$_1$+E]+[A$_1$+2B$_1$+3E], where the first and second terms correspond to the acoustic and optic phonon modes, respectively. All the optic modes are Raman (R) active. In the description below X and Y are the directions parallel to the $a$ and $b$ axes, while X' and Y' form are rotated by 45 degrees with respect to X and Y, respectively. Z is the direction parallel to the $c$ axis.

\section{Results and discussion}

\begin{figure*}[!t]
\begin{center}
\includegraphics[width=\textwidth]{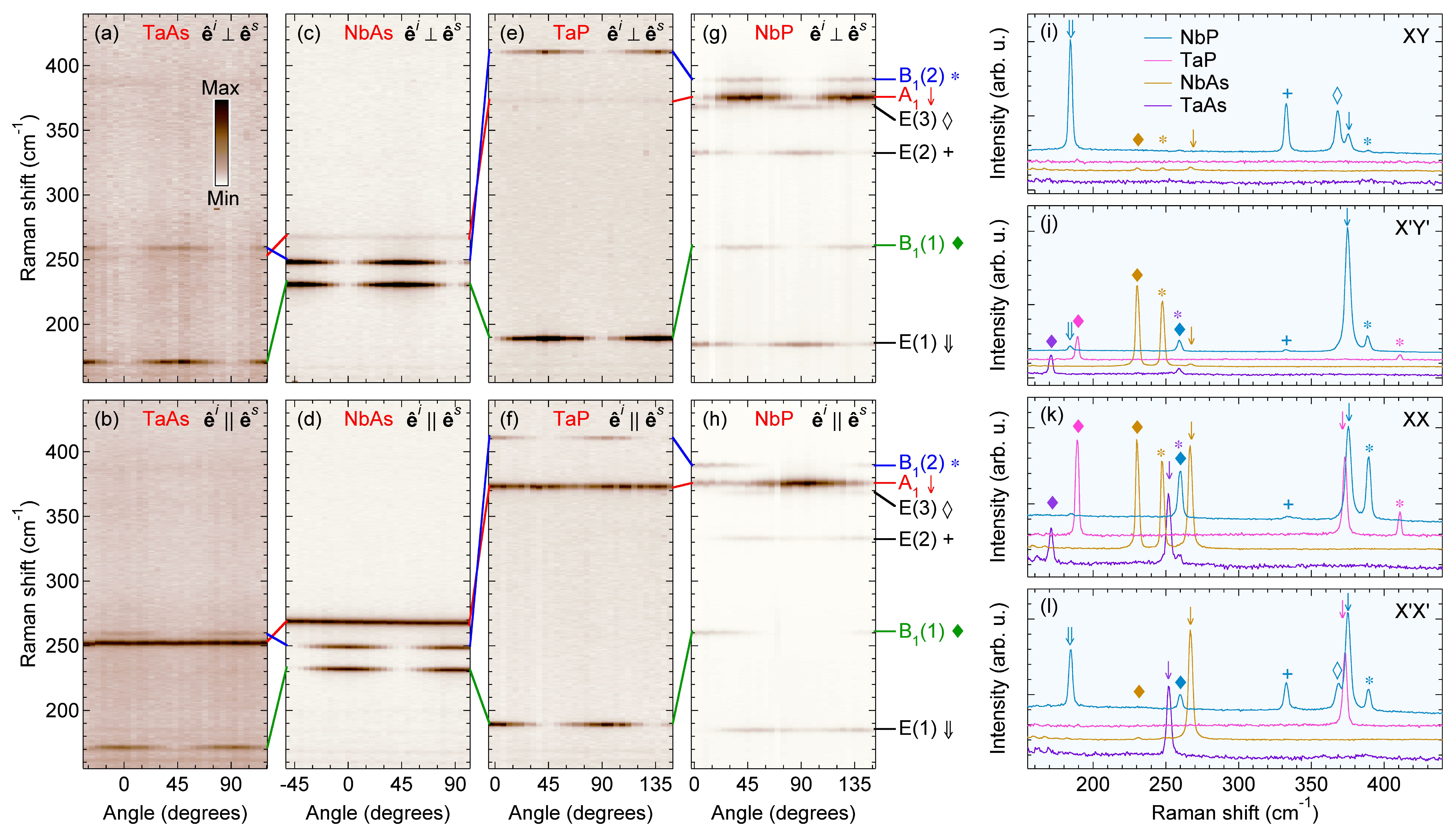}
\end{center}
\caption{\label{in_plane}(Colour online). (a), (b) In-plane angular dependence of the Raman intensity of TaAs for $\vec{\hat{e}}^{i}\perp\vec{\hat{e}}^{s}$ and $\vec{\hat{e}}^{i}||\vec{\hat{e}}^{s}$, respectively. The data on TaAs, reproduced here for completeness, are from Ref. \cite{HW_LiuPRB92}, copyright \copyright~(2015) by the American Physical Society. (c), (d) Same as (a) and (b) but for NbAs. (e), (f) Same as (a) and (b) but for TaP. (e), (f) Same as (a) and (b) but for NbP, with a larger misalignment than for the other samples. The relative colour scale for all the intensity plots is given in (a). The colour lines between the panels (a)-(h) are guides to indicate the energy shift of the modes, which are indicated on the right side of panels (g) and (h), with the corresponding colour. Each mode is associated with a symbol that helps identifying the peaks in panels (i)-(l). (i)-(l) Comparison of the Raman intensity spectra of $AX$ ($A$ = Ta, Nb; $X$ = As, P) measured in the XY, X'Y', XX and X'X' configurations of polarization, respectively. The peaks are identified by the symbols associated to each mode, with the proper colour associated to each compound, as shown in panel (i).}
\end{figure*}

In a previous work we showed that angle-dependent Raman measurements can be used to determine the symmetry of the modes observed \cite{HW_LiuPRB92}. We now apply the same procedure to NbAs, TaP and NbP. The corresponding Raman intensity plots are displayed in Fig. \ref{in_plane}, with the results on TaAs from Ref. \cite{HW_LiuPRB92} reproduced here for sake of comparison. At the exception of NbP, which shows a large misalignment due to the morphology of our sample, each material exhibits 3 peaks that can be associated to phonons probed with the incident ($\vec{\hat{e}}^{i}$) and scattered ($\vec{\hat{e}}^{s}$) light beams are parallel to the surface of the sample. The sample misalignment for our NbP sample leads to the observation of all the Raman phonons (6). As expected, peaks with the same symmetries can be observed in all materials. Hence, we observe one peak with a nearly constant intensity with respect to the in-plane angle in the $\vec{\hat{e}}^{i}||\vec{\hat{e}}^{s}$ configuration but with nearly zero intensity at 0 degree (and at every 90 degree steps) in the $\vec{\hat{e}}^{i}\perp\vec{\hat{e}}^{s}$ configuration. These peaks necessarily have the A$_1$ symmetry. Similarly, for each material there are two peaks showing zeros of intensity at 45 degrees (and at every 90 degree steps) with $\vec{\hat{e}}^{i}||\vec{\hat{e}}^{s}$, but with zeros of intensity at 0 degree (and at every 90 degree steps) for $\vec{\hat{e}}^{i}\perp\vec{\hat{e}}^{s}$. These peaks correspond to B$_1$ phonons. Finally, the three extra peaks in our measurements of NbP [Figs. \ref{in_plane}(g) and \ref{in_plane}(g)] exhibit the same angular dependence, confirming that they can be described by the same irreducible representation of point group $C_{4v}$, in this case E. 

In the simplest models, the frequency $\omega$ of a phonon mode is proportional to $1/\sqrt{M}$, where $M$ is the mass of the vibrating atom. Consequently, we expect that when comparing the same vibration mode for different $AX$ ($A$ = Ta, Nb; $X$ = As, P) compounds, the mode frequency will vary significantly only when the main atoms involved in the vibration are changed. For example, the A$_1$ mode in TaAs increases by only about 16 cm$^{-1}$ in NbAs, whereas larger shifts of more than 120 cm$^{-1}$ are found for TaP and NbP, which have approximately the same A$_1$ mode energy (see Table \ref{table_modes}). In agreement with previous first-principles calculations, we conclude experimentally from this observation that the A$_1$ mode involves mainly As or P atoms, for which the atomic masses are 74.9 u. a. and 31.0 u. a., respectively. Indeed, the ratio of the square of the average of the A$_1$ mode energy in TaP and NbP, over the square of the average of the A$_1$ mode energy in TaAs and NbAs (squared frequency average ratio), gives 2.1, which is similar to the 2.4 ratio of the As and P atomic masses. In the same way, there is only about 10 cm$^{-1}$ difference between the energy of the B$_1(2)$ mode in TaAs and NbAs, but there is an increase of more than 130 cm$^{-1}$ in the mode energy in TaP and NbP as compared to TaAs, indicating that the B$_1(2)$ phonon mode also involves mainly As or P vibrations. Here again we checked that the squared frequency average ratio is 2.5, very close to 2.4. In contrast, the similarities in the B$_1(1)$ mode energies in TaAs (171.7 cm$^{-1}$) and TaP (189.2 cm$^{-1}$) on one hand, and in NbAs (231.7 cm$^{-1}$) and NbP (259.2 cm$^{-1}$) on the other hand, indicate that the B$_1(1)$ phonon involves mainly vibrations of Ta and Nb, which have atomic masses of 180.9 u. a. and 92.9 u. a., respectively. In this case the squared frequency average ratio (NbAs-NbP over TaAs-TaP) is 1.9, which is equivalent to the ratio of the atomic masses of Ta and Nb.

\begin{table}[t!]
\begin{center}
\caption{\label{table_modes} Energy (in cm$^{-1}$) of the Raman phonon modes in TaAs, NbAs, TaP and NbP.}
\begin{tabular}{cccccc}
\hline
Mode&Main atoms&TaAs \cite{HW_LiuPRB92}&NbAs&TaP&NbP\\
\hline
A$_1$&As(P)&251.9&268.2&373.3&375.5\\
B$_1(1)$&Ta(Nb)&171.7&231.7&189.2&259.2\\
B$_1(2)$&As(P)&259.2&249.0&410.9&389.2\\
E$(1)$&Ta(Nb), As(P)&125.6&149.2&141.6&184.5\\
E$(2)$&As(P)&232.6&232.8&337.3&332.7\\
E$(3)$&As(P)&260.9&249.7&378.1&368.8\\
\hline
\end{tabular}
\begin{raggedright}
\end{raggedright}
\end{center}
\end{table}

\begin{figure}[!t]
\begin{center}
\includegraphics[width=\columnwidth]{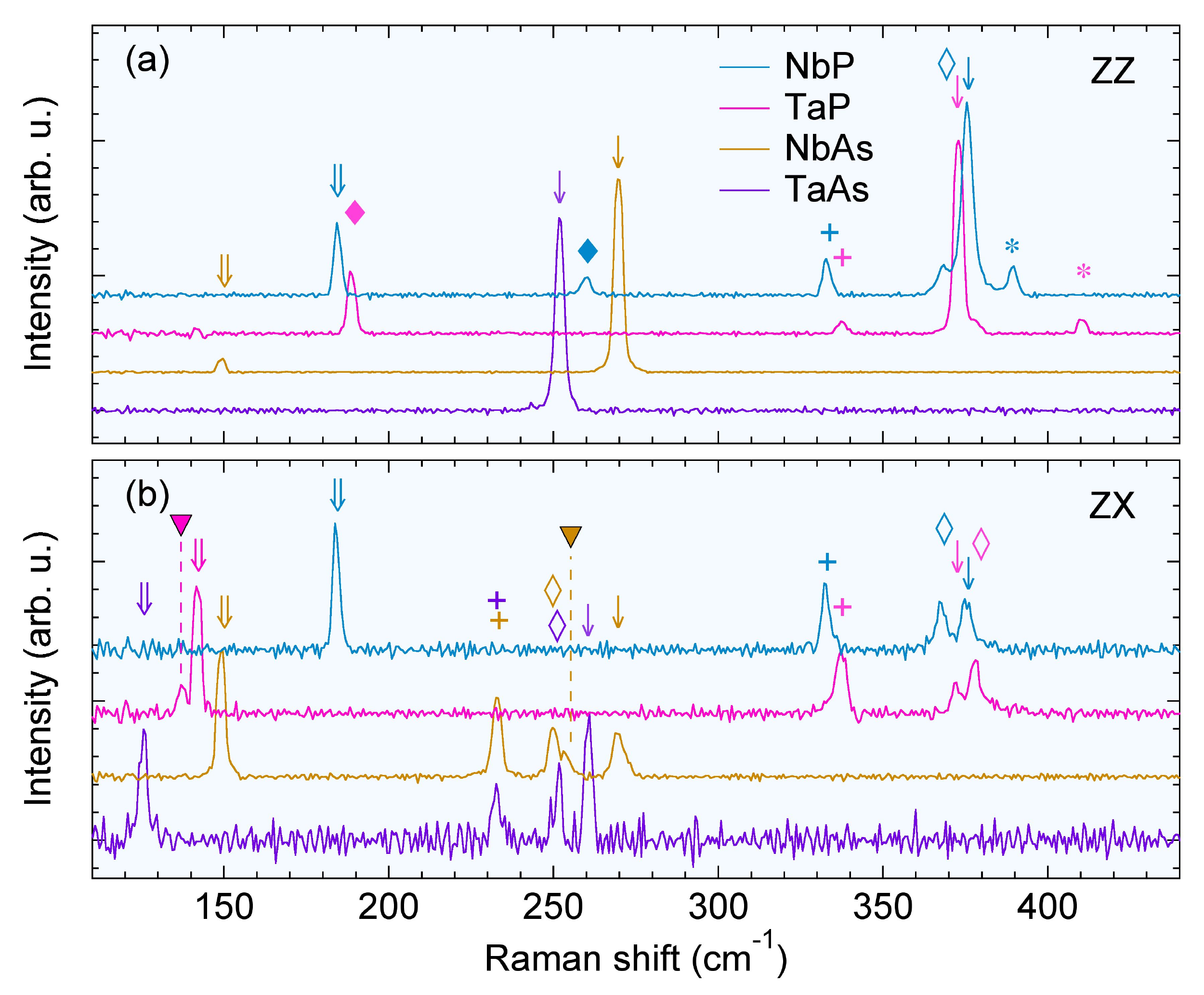}
\end{center}
\caption{\label{out_plane}(Colour online). (a) Comparison of the Raman intensity spectra of $AX$ ($A$ = Ta, Nb; $X$ = As, P) recorded with $\vec{\hat{e}}^{i}||\vec{\hat{e}}^{s}$ approximately in the ZZ configuration. The background has been removed and the spectra shifted up for better comparison. The peaks are identified by the symbols associated to each mode introduced in Fig. \ref{in_plane}, with the proper colour associated to each compound. (b) Same as (a) but for $\vec{\hat{e}}^{i}\perp\vec{\hat{e}}^{s}$ (ZX). The reverse triangles refer to extra peaks for which the origin is unclear. The data on TaAs, reproduced here for completeness, are from Ref. \cite{HW_LiuPRB92}, copyright \copyright~(2015) by the American Physical Society.}
\end{figure}

We now turn our attention to spectra recorded with the ZZ and ZX polarization configurations, which are displayed in Figs. \ref{out_plane}(a) and \ref{out_plane}(b), respectively. For perfect alignment and in the absence of defects, the E modes should be probed by the ZX configuration, for which $\vec{\hat{e}}^{i}\perp\vec{\hat{e}}^{s}$. As reported before \cite{HW_LiuPRB92}, 4 peaks are observed in the ZX spectrum of TaAs, among which one can be assigned to the A$_1$ phonon while the others correspond to the three E modes. The spectrum in the ZZ configuration helps confirming the assignment of the A$_1$ phonon in the ZX spectrum. The situation is similar in NbAs, TaP and NbP. However, one additional peak is observed at 137.0 cm$^{-1}$ in TaP, very close to the E$(1)$ mode, and at 253.7 cm$^{-1}$ in NbAs, as a shoulder of the E$(2)$ mode. Although the origin of these extra peaks remains unclear, it is worth mentioning the possibility of local modes due to non-stoichiometric defects. Indeed, a recent transmission electron microscope study reveals the existence of such defects in the $AX$ ($A$ = Ta, Nb; $X$ = As, P) series \cite{Besara_TEM}.  

As with the A$_1$ and B$_1$ modes, we can show that the energies of the E$(2)$ and E$(3)$ modes vary approximately like the inverse square-root of the mass. While the E$(2)$ mode energy is almost the same in TaAs and NbAs, the energy of the E$(2)$ mode in TaP is only 4.6 cm$^{-1}$ larger than in NbP. We thus conclude that the E$(2)$ mode involves mainly vibrations of As(P), with a squared frequency average ratio of 2.1 comparable to the ratio of the As and P masses. Similarly, we find that the E$(3)$ mode also involves mainly vibrations of As(P), also with a squared frequency average ratio of 2.1. In contrast, it is impossible to find pairs of compounds for the E$(1)$ phonon. This is easy to understand if we assume, in agreement with first-principles calculations \cite{HW_LiuPRB92}, that both Ta(Nb) and As(P) have significant displacement amplitudes in the E$(3)$ mode.

By comparing the phonon energies across the whole $AX$ ($A$ = Ta, Nb; $X$ = As, P) series, we can also comment on the effect of chemical pressure. As with applied pressure, the substitution of one atom by another from the same column of the periodic table can result in the modification of the lattice parameters. In the same way as what would be expected for the application of external pressure on NbAs, the $c$ axis parameter, the $a$ axis parameter and the volume of the unit cell follow the sequence \cite{Besara_TEM}: TaP$<$NbP$<$TaAs$<$NbAs. Since the A$_1$ mode, which involves mainly As or P, has a larger energy in NbAs than in TaAs, and a larger energy in NbP than in TaP, we conclude that chemical pressure decreases the energy of the A$_1$ mode. Surprisingly, this behaviour is opposite to the one reported on TaAs under external pressure \cite{Y_Zhou_pressure}. In contrast to the A$_1$ mode, the B$_1$ modes become harder under chemical pressure. This is also the case for the E$(3)$ mode. While the E$(2)$ mode is basically the same in TaAs and NbAs, it also exhibits an increase under chemical pressure in TaP as compared to NbP. As for the E$(1)$ mode, the notion of chemical pressure applied to the phonons is irrelevant since both kinds of atoms are involved. Although the behaviour of the B$_1$ and E modes under external pressure is unknown, our result on the A$_1$ mode indicates that external pressure and chemical pressure cannot be regarded as equivalent, suggesting that either the nature of the bondings is modified by the atomic substitution or that the compound distortion upon chemical pressure does not follow an hydrostatic compression.  

\section{Summary}

In summary, we performed a polarized Raman study of Weyl semimetals TaAs, NbAs, TaP and NbP. We identified all the optic phonon modes of these compounds. By comparing the evolution of the different modes across that series, we determined experimentally which atoms are involved in each mode. While the A$_1$, B$_1(2)$, E$(2)$ and E$(3)$ modes involve mainly vibrations of As or P, the B$_1(1)$ correspond mainly to vibrations of Ta or Nb. Both kind of atoms are involved in the E$(1)$ mode. Disregarding the E$(1)$ mode, for which the effect of external pressure on phonons cannot be determined, we also showed that all the other phonons except the A$_1$ mode become harder with external pressure. Interestingly, the behaviour of the A$_1$ mode under chemical pressure is opposite to the one reported under external pressure, indicating that both kind of pressure effect the system differently.

\section*{Acknowledgement}

This work was supported by grants from MOST (2011CBA001001, 2011CBA00102 and 2015CB921301) and NSFC (11274362, 11534005) of China.

\bibliography{biblio_long}

\end{document}